\documentclass[aps,prc,preprint,amsmath,amssymb,showpacs,preprintnumbers,superscriptaddress]{revtex4-1}
\usepackage{CJK}
\usepackage{graphicx}% Include figure files
\usepackage{dcolumn}% Align table columns on decimal point
\usepackage{bm}% bold math
\usepackage{color}
\usepackage{hyperref}% add hypertext capabilities
\allowdisplaybreaks[4]

\begin{document}
\begin{CJK*}{UTF8}{}

\title{Nuclear landscape in a mapped collective Hamiltonian from covariant density functional theory}% Force line breaks with \\
\author{Y. L. Yang \CJKfamily{gbsn} (杨一龙)}
\affiliation{State Key Laboratory of Nuclear Physics and Technology, School of Physics, Peking University, Beijing 100871, China}

\author{Y. K. Wang \CJKfamily{gbsn} (王亚坤)}
\affiliation{State Key Laboratory of Nuclear Physics and Technology, School of Physics, Peking University, Beijing 100871, China}

\author{P.~W. Zhao \CJKfamily{gbsn} (赵鹏巍)}
\email{pwzhao@pku.edu.cn}
\affiliation{State Key Laboratory of Nuclear Physics and Technology, School of Physics, Peking University, Beijing 100871, China}

\author{Z. P. Li \CJKfamily{gbsn} (李志攀)}
\email{zpliphy@swu.edu.cn}
\affiliation{School of Physical Science and Technology, Southwest University, Chongqing 400715, China}

\begin{abstract}
The nuclear landscape is investigated within the triaxial relativistic Hartree-Bogoliubov theory with the PC-PK1 density functional, and the beyond-mean-field dynamical correlation energies are taken into account by a microscopically mapped five-dimensional collective Hamiltonian without additional free parameters.
The effects of triaxial deformation and dynamical correlations on the nuclear landscape are analyzed.
The present results provide a better description of the experimental binding energies of even-even nuclei, in particular for medium and heavy mass regions, in comparison with previous global calculations with the state-of-the-art covariant density functionals DD-PC1 and TMA.
The inclusion of the dynamical correlation energies plays an important role in the PC-PK1 results.
It is emphasized that the nuclear landscape is considerably extended by the PC-PK1 functional in comparison with the previous results obtained with the covariant density functionals DD-PC1 and TMA, which may be due to the different isovector properties in the density functionals.
\end{abstract}

%\pacs{21.60.Jz, 21.10.-k, 21.10.Re, 27.20.+n}
% 21.60.Jz Nuclear Density Functional Theory and extensions
%21.10.-k Properties of nuclei; nuclear energy levels
%21.10.Re Collective levels
%21.60.Ev Collective models
%21.60.Cs   Shell model
%23.20.-g Electromagnetic transitions
%23.20.Js Multipole matrix element
%27.20.+n  6 A 19
%27.60.+j 90  A 149
%27.50.+e 59  A  89

\maketitle

%*********************************************************%
%---------------------Introduction------------------------%
%*********************************************************%
\section{Introduction}
Nuclei are self-bound systems with a certain number of protons and neutrons.
Despite enormous research effort, it is still an open question how many species of nuclei are bound in the universe.
Until now, the masses of approximately 3000 nuclei have been known from experiment~\cite{Wang2017Chin.Phys.C030003},
and the location of the proton drip line has been delineated up to protactinium ($Z = 91$).
However, the position of the neutron drip line beyond $Z = 8$ is still very elusive.
The quest for the limits of nuclear binding is closely connected to the question about the origin of elements~\cite{Burbidge1957Rev.Mod.Phys.547}.
In particular, the astrophysical rapid neutron capture process, which is responsible for the generation of many heavy elements, relies on the properties of nuclei very close to the neutron drip line, and these properties are not experimentally available in the foreseeable future.
Therefore, theoretical predictions across the nuclear landscape are required.
Nuclear density functional theory (DFT) is the most promising microscopic approach for this purpose.
It is based on a universal energy density functional, and with only a few parameters, it can describe not only the binding energies of nuclei but also other key quantities for simulating the nucleosynthesis process including $\beta$-decay rates and fission rates in a unified way~\cite{Bender2003Rev.Mod.Phys.121}.

There have been several global calculations for the nuclear chart based on both nonrelativistic and relativistic nuclear DFTs.
Axially deformed Hartree-Fock~\cite{Tajima1996Nucl.Phys.A23} and relativistic mean-field (RMF) calculations~\cite{Lalazissis1999At.DataNucl.DataTables1} were carried out for the ground-state properties of even-even nuclei.
Later, calculations for all nuclei including the odd-nucleon systems were performed with both the axially deformed Hartree-Fock approach~\cite{Tondeur2000Phys.Rev.C024308} and the RMF approach~\cite{Geng2005Prog.Theor.Phys.785}.
In all these calculations, the pairing correlations are accounted for with the Bardeen-Cooper-Schrieffer (BCS) method, which is questionable in the region near drip lines since it does not take into account the continuum properly~\cite{Meng2016}.

The axially deformed Hartree-Fock-Bogoliubov (HFB) model was employed to calculate the ground-state properties of even-even nuclei from the proton drip line to the neutron drip line in Ref.~\cite{Stoitsov2003Phys.Rev.C054312}.
A recent systematic study to estimate the limits of the nuclear landscape was performed with the axially deformed HFB model employing six sets of functionals~\cite{Erler2012Nature509}.
Similar studies in covariant DFT were reported in Ref.~\cite{Afanasjev2013Phys.Lett.B680} that employed a set of four functionals including DD-ME2~\cite{Lalazissis2005Phys.Rev.C024312}, DD-PC1~\cite{Niksic2008Phys.Rev.C34318}, DD-ME$\delta$~\cite{Roca-Maza2011Phys.Rev.C54309}, and NL3$^\ast$~\cite{Lalazissis2009Phys.Lett.B36}.
Recently, the limits of the nuclear landscape were explored using the spherical relativistic continuum Hartree-Bogoliubov (RCHB) theory~\cite{Xia2018AtomicDataandNuclearDataTables1215}, in which the continuum effects are described more properly in the coordinate space.

Although triaxiality plays an important role in describing many nuclear phenomena, such as nuclear low-lying spectra~\cite{Heyde2011Rev.Mod.Phys.1467}, nuclear chirality~\cite{Frauendorf1997Nucl.Phys.A131}, and nuclear wobbling~\cite{Bohr1975}, it was not considered in most previous studies of the nuclear landscape.
Only in Ref.~\cite{Delaroche2010Phys.Rev.C014303} was a systematic study of low-energy nuclear structure carried out using the triaxial HFB theory together with a mapped five-dimensional collective Hamiltonian (5DCH) for even-even nuclei with proton numbers $Z=10$ to $Z=110$ and neutron numbers $N \le 200$, and the Gogny D1S interaction was employed in the nuclear Hamiltonian~\cite{Decharge1980Phys.Rev.C1568}.
The 5DCH is also applied to the optimization of the Gogny D1M force, with an explicit account of the quadrupole correlation energies~\cite{Goriely2009Phys.Rev.Lett.242501, Goriely2016Eur.Phys.J.A202}.
In these works, the 5DCH is employed to extend the mean-field solutions to describe spectroscopic properties, and it also brings the beyond-mean-field dynamical correlation energies (DCEs) to nuclear ground states.
Another way to calculate the DCEs is the quantum number projection and generator coordinate method (GCM), which has been adopted to systematically study the ground-state properties of even-even nuclei by imposing axial symmetry~\cite{Bender2005Phys.Rev.Lett.102503,Bender2006Phys.Rev.C034322, Bender2008Phys.Rev.C054312, Bender2011Phys.Rev.C064319}.
However, similar studies including triaxiality so far have been too time consuming to be implemented for large-scale calculations.

The importance of DCEs in nuclear masses was been revealed in Ref.~\cite{Lu2015Phys.Rev.C027304} with the 5DCH based on triaxial RMF plus BCS calculations~\cite{Niksic2009Phys.Rev.C034303,Li2009Phys.Rev.C054301} with the PC-PK1 density functional~\cite{Zhao2010Phys.Rev.C054319}.
The root-mean-square (rms) deviation from the experimental nuclear masses is reduced significantly down to 1.14 MeV for even-even nuclei without any additional parameters beyond the density functional.
It is at the same level as those from the relativistic DD-MEB1 and DD-MEB2 functionals~\cite{PenaArteaga2016Eur.Phys.J.A320}, which are determined by large-scale fits to essentially all the experimental masses with correlation energies considered phenomenologically.

Therefore, it is interesting to explore the limits of the nuclear landscape with the triaxiality, the dynamical correlations, and the PC-PK1 density functional.
In the present work, the even-even nuclei with $8\leq Z\leq 104$ from the proton drip line to the neutron one are calculated within the triaxial relativistic Hartree-Bogoliubov (RHB) theory with PC-PK1 density functional, and the DCEs are taken into account by a microscopically mapped 5DCH~\cite{Niksic2009Phys.Rev.C034303,Li2009Phys.Rev.C054301}, which was successfully used for describing the nuclear low-lying states~\cite{Fu2013Phys.Rev.C054305,Li2012Phys.Lett.B470} and shape evolution~\cite{Xiang2012Nucl.Phys.A116,Li2013Phys.Lett.B866}.
The results for light nuclei from O ($Z=8$) to Sn ($Z=50$) were reported in Ref.~\cite{Yang2020Chin.Phys.C034102}.
The goals of the present work are (i) the systematic study of two-proton and two-neutron drip lines with the triaxial RHB framework using the PC-PK1 density functional, (ii) the analysis of triaxial and beyond-mean-field effects on the nuclear landscape, and (iii) the comparison of the drip lines obtained with the PC-PK1 density functional and other density functionals.

%==================Theoretical framework===================%
\section{Theoretical framework and numerical details}
The RHB theory provides a unified and self-consistent treatment of mean fields and pairing correlations.
The detailed formulas of the RHB theory can be seen in Refs.~\cite{Ring1996Prog.Part.Nucl.Phys.193,Vretenar2005Phys.Rep.101,Meng2006Prog.Part.Nucl.Phys.470,Niksic2011Prog.Part.Nucl.Phys.519},
and the microscopically mapped five-dimensional collective Hamiltonian with the covariant DFT can be found in Refs.~\cite{Niksic2009Phys.Rev.C034303,Li2009Phys.Rev.C054301}.
In this work, the lowest energy state in the potential energy surface provides the ground state at the mean-field level,
and diagonalization of the 5DCH yields the energy of the collective ground state $0_1^+$~\cite{Niksic2009Phys.Rev.C034303, Li2009Phys.Rev.C054301}.
The DCE $E_\mathrm{corr}$ is defined as the energy difference between the mean-field ground state and the collective $0_1^+$ state.

As mentioned above, the PC-PK1 density functional~\cite{Zhao2010Phys.Rev.C054319} is employed in the present work and, in the pairing channel, a finite-range separable pairing force with the pairing strength $G = 728$ MeV$\cdotp$fm$^3$~\cite{Tian2009Phys.Lett.B44} is adopted.
The triaxial RHB equation is solved by expanding the quasiparticle wavefunctions in terms of three-dimensional harmonic oscillator basis in Cartesian coordinates~\cite{Niksic2014Comput.Phys.Comm.1808}.
For nuclei with $Z<20$, $20\leq Z<82$, and $82\leq Z\leq104$, the harmonic oscillator basis contains respectively 12, 14, and 16 major shells, and they were examined to be able to provide converged results.

Adiabatic deformation constrained RHB calculations are performed in the full $\beta$-$\gamma$ plane $(0^\circ \leq \gamma\leq 60^\circ)$ for each nucleus.
Similar to the previous research~\cite{Lu2015Phys.Rev.C027304}, the maximum value of $\beta$ is chosen to be $\beta=1.20$ for nuclei with $Z<20$ and $\beta=0.72$ for heavier ones.
The obtained quasiparticle energies and wavefunctions at each $\beta$-$\gamma$ point are then used to calculate the collective parameters of the 5DCH and, thus, the DCE.
In the present 5DCH calculations, we only consider nuclei whose RHB states exhibit negative chemical potentials in the full $\beta$-$\gamma$ plane.
For closed-shell nuclei, the Gaussian overlap approximation adopted within the 5DCH approach may lead to erroneous unphysical correlations, and the DCE is then imposed to be zero if it occurs.

%================Results and discussion====================%
\section{Results and discussion}
\begin{figure*}[!htbp]
  \centering
  \includegraphics[width=0.9\textwidth]{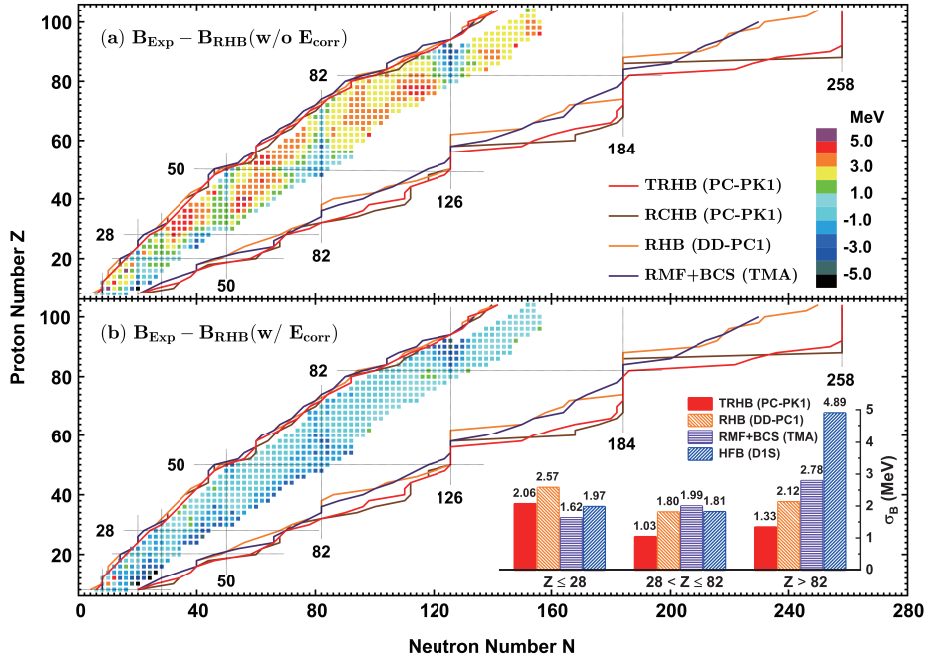}\\
  \caption{(Color online) The bound even-even nuclei obtained in the triaxial RHB calculations using PC-PK1 (a) without and (b) with the dynamical correlation energies.
  The differences between the experimental binding energies~\cite{Wang2017Chin.Phys.C030003} and the calculated ones are denoted by colors.
  The two-proton and two-neutron drip lines are also shown as solid lines in comparison with the ones obtained in the previous works: the spherical RCHB calculations with PC-PK1~\cite{Xia2018AtomicDataandNuclearDataTables1215}, the axial RHB calculations with DD-PC1~\cite{Agbemava2014Phys.Rev.C054320}, and the axial RMF plus BCS calculations with TMA~\cite{Geng2005Prog.Theor.Phys.785}.
  The inset shows the root-mean-square deviations of binding energies $\sigma_B$ in different mass regions for the triaxial RHB results with PC-PK1, in comparison
   with those for the axial RHB results with DD-PC1~\cite{Agbemava2014Phys.Rev.C054320}, the RMF plus BCS results with TMA~\cite{Geng2005Prog.Theor.Phys.785}, and the nonrelativistic triaxial HFB results with D1S~\cite{Delaroche2010Phys.Rev.C014303}.
  The dynamical correlation energies are included in the PC-PK1 and D1S results.}\label{fig1}
\end{figure*}

In Fig.~\ref{fig1}, the bound even-even nuclei obtained in the triaxial RHB calculations with PC-PK1 are depicted in comparison with those from the spherical RCHB calculations with PC-PK1~\cite{Xia2018AtomicDataandNuclearDataTables1215}, the axial RHB calculations with DD-PC1~\cite{Agbemava2014Phys.Rev.C054320}, and the axial RMF plus BCS calculations with TMA \cite{Geng2005Prog.Theor.Phys.785}.
In the present work, nuclei are considered bound if the two-nucleon separation energies are positive, and in terms of the chemical potential, if the proton and neutron chemical
potentials are negative, in the mean-field calculations.

The two-proton drip line predicted by the triaxial RHB calculations with PC-PK1 is very close to the predictions of the previous calculations.
This was expected because there are many data available close to the proton drip line constraining the density functionals.
For the two-neutron drip line, however, the present PC-PK1 prediction is in general more extended than those given by other density functionals.
Such an extension of the drip line should not be caused by the triaxial effects, since almost all nuclei near the two-neutron drip line are spherical or  axially symmetric (see Fig.~\ref{fig3}).
It should stem mainly from the different isovector properties of the density functionals.

Note that the symmetry energy $J$ and its slope $L$ at the saturation density given by PC-PK1 are 35.6 and 113 MeV~\cite{Zhao2010Phys.Rev.C054319}, respectively.
The predicted neutron skin thickness of $^{208}$Pb is 0.26 fm~\cite{Zhao2010Phys.Rev.C054319} and the predicted maximum mass of neutron stars is 2.3 $M_\odot$~\cite{Sun2019AIPConferenceProceedings020020}.
These predictions are all compatible with the corresponding empirical constraints.
Nevertheless, at the present stage, one must be careful extrapolating for unknown nuclei with a given density functional determined by experimentally known nuclei.
This is because the isovector channel is not well constrained yet by the current experimental data~\cite{Piekarewicz2021Phys.Rev.C024329}.

With the same PC-PK1 functional, the positions of the neutron drip line given by the spherical RCHB calculations and the triaxial RHB one are significantly different, in particular for the medium and heavy mass regions.
In the spherical RCHB calculations, the last bound neutron-rich nucleus for each isotopic chain jumps, in many cases, from one closed shell or subshell to another, such as $N=$ 112, 126, 184, and 258.
For the triaxial RHB results, however, the neutron numbers of drip-line nuclei vary relatively smoother for adjacent isotopic chains, so the last bound neutron-rich nuclei for most isotopic chains are open-shell nuclei.
This should be due to the fact that the deformation effects provide more correlations for open-shell nuclei and, thus, make them relatively easier to be bound.

The differences between the experimental binding energies from AME2016~\cite{Wang2017Chin.Phys.C030003} and the calculated ones are also depicted in Fig.~\ref{fig1} with colors.
The results with and without the DCEs are respectively shown in the lower and upper panels.
Without the DCEs, the binding energies for most nuclei are underestimated by roughly 3 MeV.
By considering the DCEs, the calculated results and the data are in quite good agreement, and the energy deviations for most nuclei are within $\pm 1$ MeV.

\begin{table}[!htbp]
  \centering
  \caption{The root-mean-square deviations of the binding energies and two-neutron and two-proton separation energies $\sigma_B$, $\sigma_{S_{2n}}$, and $\sigma_{S_{2p}}$ (in MeV) for the triaxial RHB (TRHB) calculations with PC-PK1 with respect to the data from AME2016~\cite{Wang2017Chin.Phys.C030003}.
The results of the axial RHB calculations with DD-PC1~\cite{Agbemava2014Phys.Rev.C054320}, the axial RMF plus BCS calculations with TMA~\cite{Geng2005Prog.Theor.Phys.785}, and the nonrelativistic triaxial HFB calculations with D1S~\cite{Delaroche2010Phys.Rev.C014303} and D1M~\cite{Goriely2009Phys.Rev.Lett.242501} are also listed for comparison.
}\label{tab1}
    \begin{tabular}{lccc}
    \hline
    DFTs                                                                           & $\sigma_{B}$  & $\sigma_{S_{2n}}$  & $\sigma_{S_{2p}}$  \\ \hline
    TRHB with $E_\mathrm{corr}$ (PC-PK1)                                           & 1.31          & 0.78               & 0.80               \\
    TRHB without $E_\mathrm{corr}$ (PC-PK1)                                        & 2.64          & 1.06               & 0.93               \\
    RHB (DD-PC1)~\cite{Agbemava2014Phys.Rev.C054320}                               & 2.00          & 1.10               & 0.95               \\
    RMF+BCS (TMA)~\cite{Geng2005Prog.Theor.Phys.785}                               & 2.11          & 0.98               & 1.17               \\
    HFB with $E_\mathrm{corr}$ (D1S/D1M)~\cite{Delaroche2010Phys.Rev.C014303, Goriely2009Phys.Rev.Lett.242501}   & 2.72/0.87          & 0.79/0.74               & 0.81/0.69               \\
    HFB without $E_\mathrm{corr}$ (D1S/D1M)~\cite{Delaroche2010Phys.Rev.C014303, Goriely2009Phys.Rev.Lett.242501}& 4.75/3.31          & 1.12/0.99               & 0.95/0.86               \\
    \hline
  \end{tabular}
\end{table}

For a quantitative comparison with the previous works, the rms deviations of the binding energies ($\sigma_B$) as well as the two-neutron ($\sigma_{S_{2n}}$) and two-proton ($\sigma_{S_{2p}}$) separation energies for the present calculations are listed in Table~\ref{tab1}, in comparison with the previous DFT predictions with respect to the data available from AME2016~\cite{Wang2017Chin.Phys.C030003}.
The present triaxial RHB calculations including the DCEs with PC-PK1 significantly improve the description of the binding energies in comparison with other relativistic density functionals and also the Gogny D1S, while the D1M provides even smaller rms deviations after including the DCEs.
It should be noted that the D1M was fitted to almost all experimental masses by excluding the DCEs~\cite{Goriely2009Phys.Rev.Lett.242501, Goriely2016Eur.Phys.J.A202}, leading to quite low values for the slope of the symmetry energy $L=24.8$ MeV and neutron skin thicknesses~\cite{Sellahewa2014Phys.Rev.C054327, Inakura2015Phys.Rev.C064302}.

Without the DCEs, the rms deviation of the binding energies given by the present triaxial RHB calculations with PC-PK1 is 2.64 MeV, which is on the same level as the previous results~\cite{Geng2005Prog.Theor.Phys.785, Agbemava2014Phys.Rev.C054320} with the state-of-the-art relativistic density functionals DD-PC1~\cite{Niksic2008Phys.Rev.C34318} and TMA~\cite{Sugahara1994Nucl.Phys.A557}.
The accuracy of the PC-PK1 calculations is significantly improved by including the DCEs.
Note that the PC-PK1 functional was adjusted to the masses of 60 spherical nuclei, whose DCEs usually vanish, and a neat balance between the mass description of spherical nuclei and deformed ones was achieved.
As a result, the mean-field predictions for deformed nuclei are systematically underbound with an excellent isospin dependence~\cite{Zhao2010Phys.Rev.C054319} and, thus, the inclusion of DCEs improves the description.
This may not be the case, however, for other relativistic density functionals (for example, see Fig. 9 in Ref.~\cite{Zhao2010Phys.Rev.C054319}).
For the DD-PC1 calculations~\cite{Niksic2008Phys.Rev.C34318}, the masses of well-deformed nuclei were adopted in the fitting procedure and, thus, the underlying dynamical correlations should have already been effectively considered.
For the TMA calculations~\cite{Geng2005Prog.Theor.Phys.785}, the binding energies for most nuclei, especially for $Z\geq50$, are overestimated at the mean-field level, but dynamical correlations would provide additional binding for nuclei and, thus, worsen the overall description.
For the nonrelativistic HFB calculations with D1S~\cite{Delaroche2010Phys.Rev.C014303}, the inclusion of DCEs improves the description of the binding energies, whereas the rms deviation is considerably larger than the ones given by the relativistic density functionals.

\begin{figure*}[!htbp]
  \centering
  % Requires \usepackage{graphicx}
  \includegraphics[width=0.9\textwidth]{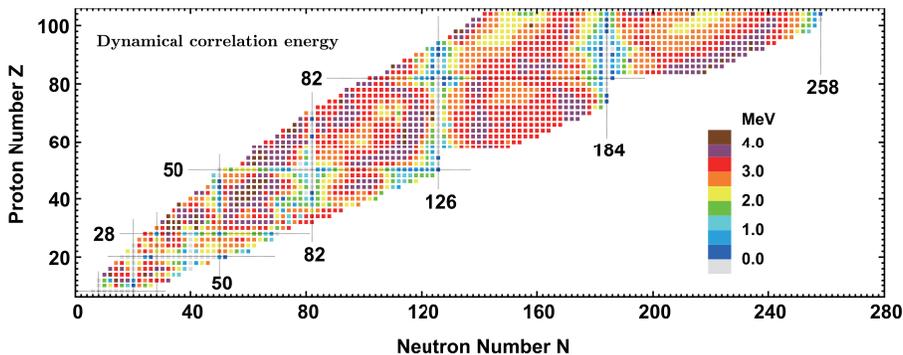}\\
  \caption{(Color online) Dynamical correlation energies calculated with the 5DCH based on the triaxial RHB calculations with PC-PK1.}\label{fig2}
\end{figure*}

The two-neutron and two-proton separation energies are described with an accuracy of around 1 MeV for all the density functionals on the mean-field level.
In the present calculations, the inclusion of DCEs significantly improves the description of two-nucleon separation energies around shell closures.
However, for most open-shell nuclei, the DCEs are usually similar for neighboring nuclei.
Therefore, the final $\sigma_{S_{2n}}$ and $\sigma_{S_{2p}}$ are only slightly improved.

It should be mentioned that the effects of dynamical correlations on many other physical aspects have been studied, including the quenching of shell closures~\cite{Bender2008Phys.Rev.C054312}, the proton-neutron interactions $\delta V_{pn}$~\cite{Bender2011Phys.Rev.C064319}, the $\beta$-decay $Q$-values~\cite{Goriely2016Eur.Phys.J.A202}, and the collective low-lying levels \cite{Delaroche2010Phys.Rev.C014303}.
Studies of these effects in the present covariant-DFT-based 5DCH method are in progress.

In the inset of Fig.~\ref{fig1}, the rms deviations of binding energies $\sigma_B$ in the light ($Z\leq28$), medium ($28<Z\leq82$) and heavy ($Z>82$) mass regions are shown for the present triaxial RHB results with PC-PK1 in comparison with previous works~\cite{Geng2005Prog.Theor.Phys.785,Delaroche2010Phys.Rev.C014303,Agbemava2014Phys.Rev.C054320}.
For the light nuclei with $Z\leq28$, the rms deviations for all calculations are around 2 MeV.
The DD-PC1 functional are relatively worse because the functional DD-PC1 was mainly fitted to deformed nuclei in medium and heavy mass regions.
The TMA results are relatively better, and this may be due to the fact that many light nuclei are included in the fitting procedure of TMA via an $A$-dependent recipe of the parameters.
For both medium and heavy mass regions, however, the present triaxial RHB results with PC-PK1 are in much better agreement with the data, in comparison with the results given by other density functionals.
Such good global performance of the PC-PK1 functional for nuclear binding energies was also seen in previous studies~\cite{Lu2015Phys.Rev.C027304, Meng2013Front.Phys.55}, where the pairing correlations are treated with the BCS approach.

In Fig.~\ref{fig2}, the DCEs calculated by the 5DCH based on the triaxial RHB calculations are shown.
They range from 2.0 to 4.0 MeV for most nuclei, and this range is similar to the previous results of 5DCH~\cite{Goriely2016Eur.Phys.J.A202} and GCM~\cite{Bender2006Phys.Rev.C034322} calculations based on nonrelativistic DFTs.
In general, the DCEs show weak dependence on nuclear masses.
Vanishing DCEs are obtained mainly for magic nuclei, and the largest ones are obtained for midshell nuclei.
Some nuclei with a single magic number, for example, Ni, Sn, and Pb isotopes, exhibit remarkable correlation energies, similar to those in Ref.~\cite{Bender2006Phys.Rev.C034322}.
Even though most of them are spherical on the mean-field level, their potential energy surfaces are either soft or exhibit a coexistence of spherical and deformed shapes.
Therefore, one expects that the DCEs for these nuclei are mainly from shape fluctuations.
The DCEs are greatly affected by the quadrupole deformations of nuclei (see Fig.~\ref{fig3}).
For nuclei with substantial deformation, e.g., the rare-earth and actinide ones centered respectively at $Z = 60, N = 100$ and $Z = 90, N = 150$, the dynamical correlations are significant because of the rotational motion.
Nevertheless, for nuclei with relatively modest deformation, e.g., the transitional nuclei located in the vicinity of the shell closures, their DCEs can be as large as those of the well-deformed nuclei, and this should be related to the pronounced vibrational motion due to the soft potential energy surfaces.

The systematics of the DCEs in the present work are similar to those obtained in the previous work with the triaxial RMF plus BCS calculations~\cite{Lu2015Phys.Rev.C027304}, while the magnitudes are generally larger by a few hundred keV.
To understand this phenomenon, taking the nucleus $^{112}$Ru as an example, we compared the pairing energies and the zero-point energies in both calculations.
As shown in the Supplemental Material~\cite{Supp}, the pairing correlations given by the Bogoliubov approach are generally stronger than those from the BCS approach; similar to previous work~\cite{Xiang2013Phys.Rev.C057301}.
The stronger pairing correlations lead to larger zero-point energies and lower collective potentials in the corresponding 5DCH and, thus, larger DCEs.

\begin{figure*}[!htbp]
  \centering
  % Requires \usepackage{graphicx}
  \includegraphics[width=0.9\textwidth]{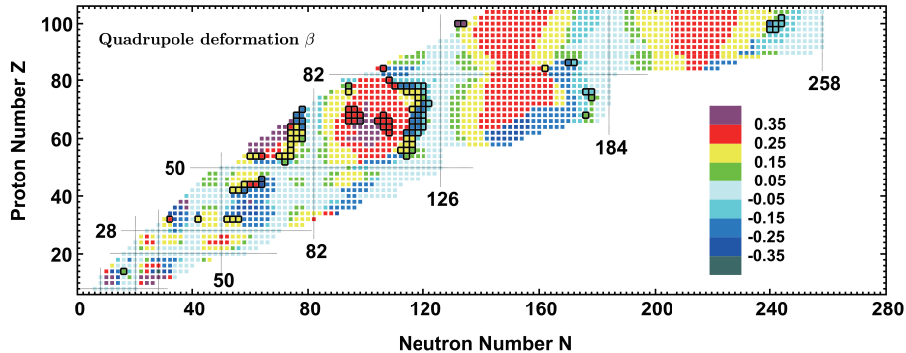}\\
  \caption{(Color online) Quadrupole deformations $\beta$ obtained in the present triaxial RHB calculations with PC-PK1.
  Triaxially deformed nuclei are denoted by black boxes.}\label{fig3}
\end{figure*}

In Fig.~\ref{fig3}, the quadrupole deformations $\beta$ given by the triaxial RHB calculations with PC-PK1 are presented.
The sign of $\beta$ is taken as positive for prolate-like nuclei $(0^\circ\leq\gamma<30^\circ)$ and negative for oblate-like ones $(30^\circ\leq\gamma\leq60^\circ)$.
Most nuclei are deformed, except for those near the closed shells.
Triaxially deformed nuclei are denoted by black boxes in Fig.~\ref{fig3}.
Most nuclei on the neutron-rich side are predicted to be spherical or axially deformed and, therefore, the neutron drip line is not influenced by the triaxiality.
In the present work, 107 even-even nuclei are predicted to be triaxially deformed.
This is fewer than the predictions by the macroscopic-microscopic finite-range liquid-drop model calculations~\cite{Moeller2006Phys.Rev.Lett.162502, Moeller2008At.DataNucl.DataTables758} and the Skyrme calculations~\cite{Scamps2021arXiv}.
The predicted triaxial regions near $N=60, Z=44$ and $N=76,Z=60$ are consistent with previous studies~\cite{Moeller2006Phys.Rev.Lett.162502,Moeller2008At.DataNucl.DataTables758,Scamps2021arXiv}.
In addition, our calculations also suggest remarkable triaxial deformations in the region with $N = 114$-120 and $Z = 54$-78, and this is similar to the results in Ref.~\cite{Scamps2021arXiv}.

%===============SUMMARY ==============
\section{Summary}

In summary, the even-even nuclei with $8\leq Z\leq 104$ are calculated from the two-proton drip line to the two-neutron one within the triaxial relativistic Hartree-Bogoliubov theory using the PC-PK1 density functional, and the beyond-mean-field dynamical correlation energies are taken into account by solving a microscopically mapped five-dimensional collective Hamiltonian without additional free parameters.
In comparison with the previous global calculations with DD-PC1 and TMA, the present PC-PK1 results provide a better description of the experimental binding energies of even-even nuclei, in particular for medium and heavy mass regions.
Triaxial deformation effects are found be small on the mean-field level, but they play an important role in the beyond-mean-field calculations for the dynamical correlation energies.
The description of the experimental binding energies is significantly improved by including the dynamical correlation energies.
Moreover, it is interesting to note that the two-neutron drip line obtained with PC-PK1 is more extended than the ones given by the covariant density functionals DD-PC1 and TMA, due to the different isovector properties rooted in the density functionals.

The data set of the calculated properties of even-even nuclei in the present calculations is provided as Supplemental Material with this article at Ref.~\cite{Supp}.

\begin{acknowledgments}
This work was partly supported by the National Key R\&D Program of China (Contracts No. 2018YFA0404400 and No. 2017YFE0116700),
the National Natural Science Foundation of China (Grants No. 11621131001, No. 11875075, No. 11935003, No. 11975031, and No. 11875225), the China Postdoctoral Science Foundation (2020M680183), and the High-performance Computing Platform of Peking University.
ZPL also acknowledges the support by the Fok Ying-Tong Education Foundation, China.
\end{acknowledgments}

%\bibliography{Reply}
%merlin.mbs apsrev4-1.bst 2010-07-25 4.21a (PWD, AO, DPC) hacked
%Control: key (0)
%Control: author (72) initials jnrlst
%Control: editor formatted (1) identically to author
%Control: production of article title (-1) disabled
%Control: page (0) single
%Control: year (1) truncated
%Control: production of eprint (0) enabled
%

\end{CJK*}
\end{document}